\documentclass[12pt, draftclsnofoot, onecolumn]{IEEEtran}
\usepackage{enumerate}
\usepackage{graphicx,indentfirst}
\usepackage{booktabs}
\usepackage{indentfirst}
\usepackage{amsmath}
\usepackage{cite}
\usepackage{stfloats}
\usepackage{multicol}
\usepackage{makecell}
\usepackage{multirow}
\usepackage{subfig}
\usepackage[english]{babel}
\usepackage{amsthm}
\usepackage{bm}
\usepackage{amsmath}
\usepackage{multirow}
\usepackage{paralist}
\usepackage{algorithm, algorithmic}
\usepackage{amssymb}
\usepackage{amssymb}
\usepackage{mathrsfs}
\usepackage{amsfonts}
\usepackage{color}
\usepackage{tablefootnote}

\theoremstyle{plain}

\IEEEoverridecommandlockouts

\begin{document}
\bibliographystyle{IEEE2}

\title{Market-Oriented Information Trading in Internet of Things (IoT) for Smart Cities}
\author{Yang~Zhang,~\IEEEmembership{Member,~IEEE,} Zehui~Xiong,~\IEEEmembership{Student Member,~IEEE,} Dusit~Niyato,~\IEEEmembership{Fellow,~IEEE,} Ping~Wang,~\IEEEmembership{Senior Member,~IEEE,} and~Zhu~Han,~\IEEEmembership{Fellow,~IEEE}\\
\thanks{Yang Zhang is with School of Computer Science and Technology, Wuhan University of Technology, China. Zehui Xiong, Dusit Niyato and Ping Wang are with School of Computer Science and Engineering, Nanyang Technological University, Singapore. Zhu Han is with the University of Houston, Houston, TX 77004 USA, and also with the Department of Computer Science and Engineering, Kyung Hee University, Seoul, South Korea.}\vspace*{-4mm}}
\maketitle

\begin{abstract}
Internet of Things (IoT) technology is a fundamental infrastructure for information transmission and integration in smart city implementations to achieve effective fine-grained city management, efficient operations, and improved life quality of citizens. Smart-city IoT systems usually involve high volume and variety of information circulation. The information lifecycle also involves many parties, stakeholders, and entities such as individuals, businesses, and government agencies with their own objectives which needed to be incentivized properly. As such, recent studies have modeled smart-city IoT systems as a market, where information is treated as a commodity by market participants. In this work, we first present a general information-centric system architecture to analyze smart-city IoT systems. We then discuss features of market-oriented approaches in IoT, including market incentive, IoT service pattern, information freshness, and social impacts. System design chanlenges and related work are also reviewed. Finally, we optimize information trading in smart-city IoT systems, considering direct and indirect network externalities in a social domain. 
\end{abstract}

\begin{IEEEkeywords}
Internet of Things, smart city, information trading, Stackelberg game, market model.
\end{IEEEkeywords}
\newpage

\section{Introduction}

Smart city~\cite{bohli.sigcomm.2009, dustdar.value.2016, mohanty.everythingiot.2016} applications aim to improve quality of lives of citizens, increase efficient public and private resource utilization, and reduce pollution, nuisance, and crime. Smart city introduces a unique requirement to Internet of Things (IoT) system designs. Specifically, IoT systems will be used by numerous and diverse smart-city applications and users with large quantities of data generated. This is significantly different from general IoT systems that are designed to support specific single-purpose applications. For example, GPS trace data collected from commuters' smartphones and video images from city cameras can be jointly used by a government in city planning, public transportation operators in allocating and routing buses and trains, logistic businesses in optimizing package delivery, and the commuters themselves in trip planning. As such, information collected and used in the IoT systems can be treated as a commodity which can be traded among information producers, processors, sellers, and customers/users. Moreover, as smart-city applications mainly aim at ordinary users, e.g., governments and individual users, it is unnecessary to reveal too many IoT device layer details in the applications, such as structures and functionalities. Processed data and services are usually required to be delivered to users. Some practical scenarios and examples of IoT for smart-city applications are shown in Table~\ref{table:iotsc.examples}.

In this work, we mainly discuss a market-oriented vision of smart-city IoT systems with emerging information exchange and resource allocation techniques. In Section~\ref{Sec:Intro}, we present an information-centric layered architecture of smart-city Iot applications. We identify some unique features of a socio-technical paradigm that makes smart-city IoT systems different from conventional IoT from a ``value of information'' perspective in Section~\ref{Sec:Market}. Typical design considerations with related work are also reviewed in Section~\ref{Sec:CaseStudy}. Next, in Section~\ref{Sec:ProposedGame}, we propose a game-theoretic market model for information trading in smart-city IoT scenarios. Some numerical studies that show the smart-city stakeholders' behaviors and benefits of the information trading are highlighted.

		\begin{small}
		\begin{table}[!tbp]
		\caption{IoT for Smart Cities: Examples.}
		\label{table:iotsc.examples}
		\centering
		\begin{tabular}{p{1.8cm}||p{2.4cm}|p{3.4cm}|p{3.4cm}|p{3.4cm}}
		\hline
		\bf{Type} & \bf{Application} & \bf{Sensor/Platform} & \bf{Data/Business Model} & \bf{Service} \\
		\hline
		\hline
		
		Industry & Miami-Dade water processing\tablefootnote{https://customers.microsoft.com/en-us/story/miami-dade-water-and-sewer-government-azure-sql-database-azure-iot-suite-sql-server-2016-power-bi-en } & Sensors for pressure, flow rates, rainfall, and more / Platform for quick collected data processing & Data stored and processed at a data warehouse for local water and sewer department to manage water supply & Real-time services and reactions to provide services to residents \\
		\hline
		
		Infrastructure & Smart Nation Singapore: Building environment\tablefootnote{http://www.iotjournal.com/articles/view?15220/3 } & In-building air conditioning sensors and control system & Monitoring energy consumption and control airconditioner behaviors & Free-of-charge deployment of IoT applications, customer payments based on energy saved by deploying the application \\
		\hline
		
		Environment & CPCB Gange river cleaning\tablefootnote{https://blogs.microsoft.com/iot/2017/11/21/cleaning-ganges-river-help-iot } & Water quality sensors, including fluoride levels, temperature, color & Capturing real-time pollution data of the river & Providing data to the government for pollution control and trend analyses. Data stored in data warehouses \\
		\hline
		
		City Management & Padova city\tablefootnote{http://hit.psy.unipd.it/padova-smart-city } & Sensors monitoring environment, e.g., pollution, public lighting, human behavior & Layered architecture. Information-centric, data processed in several phases for city decisions & Providing different levels of information, applications and value-added services for the city \\
		\hline
		
		Healthcare & Saensuk Smart City: Elderly care\tablefootnote{http://www.dell.com/learn/us/en/uscorp1/press-releases/2016-07-26-saensuk-smart-city-pilots-first-healthcare-iot-project-with-dell-intel } & Cloud-based end-to-end platform & Data collection at home by IoT Gateways, processed by nursing cloud in real time & Providing local nursing services to the elderly \\
		\hline
		
		Service & Nokia S2aaS\tablefootnote{https://www.nasdaq.com/article/nokia-launches-blockchain-powered-iot-sensing-as-a-service-for-smart-cities-cm925935 } & Blockchain-based built-in micropayment platform & Smart contracts for micro-transactions & Sensing-as-a-Service in information providing \\
		\hline
		
		Business & Smart Dubai blockchain strategy\tablefootnote{https://www.ibm.com/blogs/blockchain/2017/04/blockchain-in-dubai-smart-cities-from-concept-to-reality } & Blockchain records & Healthcare records, certificates, ID verifications, financial exchanges, and so on. & Government and market services provided to business users, residents and tourists  \\
		
		\hline
		\end{tabular}
		\end{table}
		\end{small}
		
\section{IoT Enabled Smart City}		\label{Sec:Intro}
	\subsection{IoT and Smart City}
	In conventional IoT applications, e.g., wireless sensor networks (WSNs), technical problems such as physical infrastructure, network architecture and communication techniques have been relatively well studied. However, as human and social factors are heavily involved in smart city applications, in recent studies, it is more suitable to consider smart city as a socio-technical system where the relationships among physical/virtual devices and social participants are driven by values and incentives. As such, market-oriented approaches can be employed as mechanisms to implement practical smart-city IoT systems. 

	\subsection{IoT for Smart City: An Information-centric Architecture}
	Evidences have shown that smart-city IoT applications are data intensive~\cite{mulligan.bizmodel.2013, mohanty.everythingiot.2016, baowei.mswim.2017} with the following features:
	\begin{itemize}
		\item {\bf Information quantity}: A large amount of data and information need to be generated and transported in smart-city IoT applications.
		
		\item {\bf Information heterogeneity}: Smart-city IoT systems collect and use various and diverse data from numerous sources with a variety of types, formats, and attributes. Moreover, the role of each participant changes regularly. These data is used jointly to achieve a certain goal. 
		
		\item {\bf Information quality}: Information exchanged in smart-city IoT systems may have different inherent quality. For example, data collected from an advanced smartphone typically has higher accuracy than that from a small sensor. Additionally, ``age of information'' is another important quality metric. The information which is promptly retrieved has higher quality than that which is delayed.
	\end{itemize}
	Consequently, instead of a network-centric architecture focusing on end-to-end information transmissions, an information-centric architecture becomes more promising for the design and deployment of smart-city IoT systems.

		\begin{figure}[htbp]
		\begin{center}
		\includegraphics[width=0.90\textwidth]{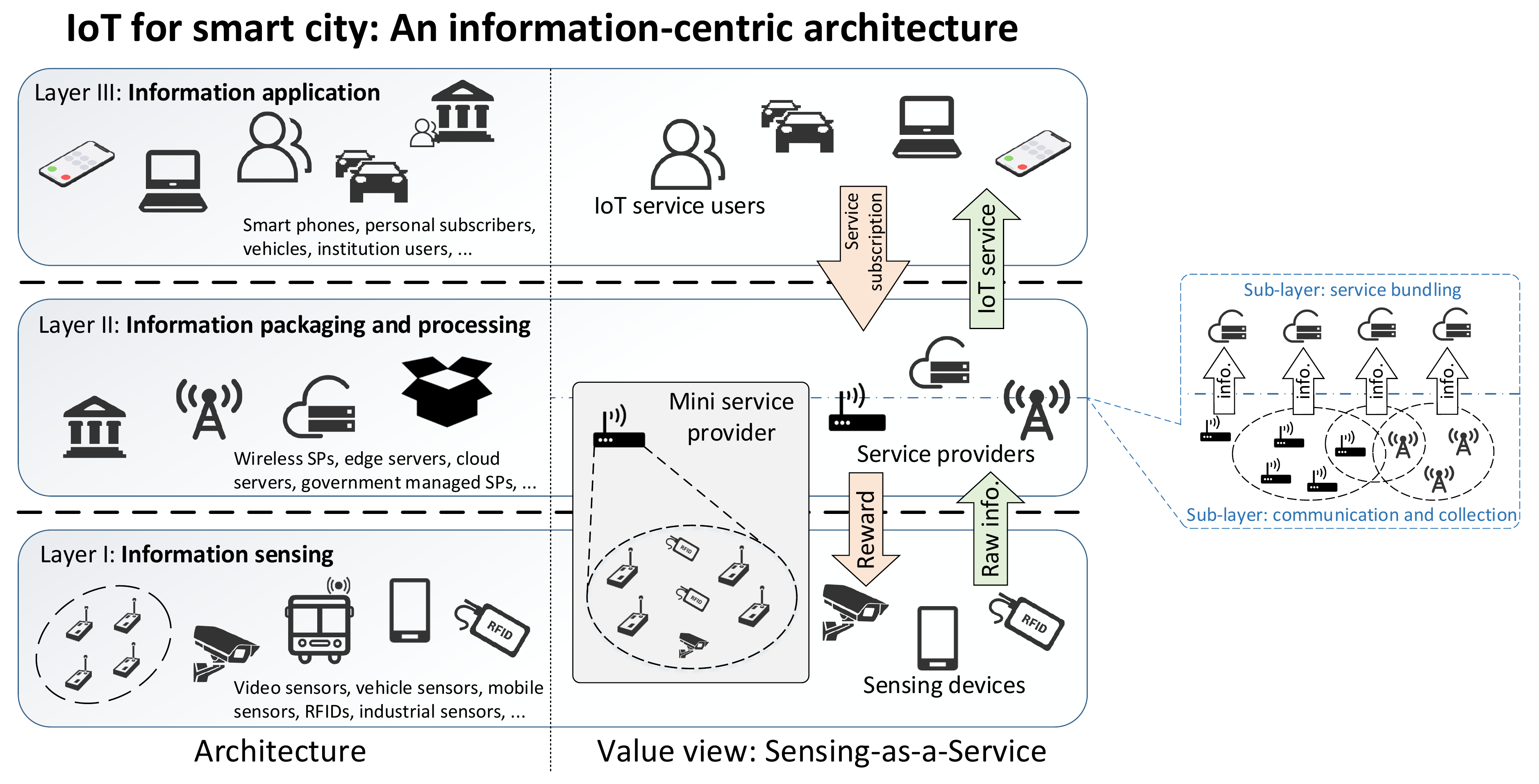}
		\caption{\small Information-centric architecture of IoT for smart city systems.}
		\label{Fig:Arch}
		\end{center}
		\end{figure}
	
	An information-centric reference system architecture for an IoT smart city is shown in Fig.~\ref{Fig:Arch}. The architecture has three layers with different functionality.
	\begin{itemize}
		\item{\bf Layer I, information sensing and generating layer}: Components working in this layer can sense, collect and generate ``raw'' data either from the environment or within the network devices, e.g., surveillance records of city video cameras. The layer provides interfaces at the edge of smart-city IoT architecture and physical world.
		
		\item{\bf Layer II, information packaging and processing layer}: Raw data from Layer I is processed in this layer and used to provide services. The services can be formed as structured and interpreted information bundles by selectively combining and employing the raw data passed to this layer, e.g., using machine learning algorithms.
		
		\item{\bf Layer III, information application layer}: Information services in this layer are applied to serve the needs and demand of city users. For example, traffic forecast information services can be adopted by smart transportation users to predict traffic situation and plan their trips.
	\end{itemize}

The information-centric smart-city IoT architecture allows cross layer implementations, and the key component is an information flow rather than device placement as in conventional IoT systems. This information-centric architecture promotes cross-layer designs and implementations. As many devices in smart-city systems have more computing capability, e.g., smartphones, primitive functions can be provided locally by each device (or a group of devices) to meet immediate information requirements of smart-city users. For example, noise can be removed from the sensed information by sophisticated algorithms running by the smartphones. In this regard, IoT devices defined as micro providers~\cite{bohli.sigcomm.2009} can operate across Layers I and II, as shown in Fig.~\ref{Fig:Arch}. The information-centric architecture can be employed in market-oriented modeling and analysis for information trading in smart-city IoT systems. With this architecture, entities in smart city can only concentrate on the content and timeliness while acquiring data from the system, instead of low-level communication issues.

\section{Market-oriented IoT Information Trading for Smart Cities}	\label{Sec:Market}

In this section, we present smart-city IoT systems from the ``value of information and services'' perspective. Information is treated as goods, and information transfer is considered to be market transactions. We propose the definition of market-oriented approaches for information trading in smart-city IoT systems. Some key components and mechanisms in a smart-city information market are discussed.

	\subsection{IoT for Smart City: A Value of Information and Services}
	A smart city is not a simple combination of sensors and sensor users. As a framework of social governance and business solution, technological and societal entities are all adopted in smart city IoT implementations as the core components and information sources. In particular, a smart city involves human participants in forms of individuals and social groups. In this regard, information generated in smart city systems has its inherent values to owners and other participants. 	Activities to handle and process the collected information can be seen as value-adding or value generation processes. Consequently, information becomes expensive in value when the scale of data sensing, collection and processing as well as the complexity of the smart city applications increase. Therefore, transfers of raw data and processed information should involve value interpretation/proposition and reward implication in terms of either monetary or other types of incentive as in market trading.

	One of the incentive schemes in real smart-city IoT applications is the Travel Smart Rewards (TSR) program of Singapore\footnote{https://www.travelsmartrewards.lta.gov.sg/}. In the program, public transit card holders/users, i.e., the information sources, can register and share their travel records with the Land Transport Authority (LTA) of Singapore, i.e., the information processor, for research and public transit management uses, i.e., value-added services. Upon sharing the information, a cash reward is paid by LTA to the registered users. LTA can use such information to schedule trains and buses to reduce fuel consumption, improve waiting time, and driver and operator management, hence increasing operating efficiency and profit.
	
	Evidently, smart-city IoT applications are more concerned about ``value of information'' than classical performance metrics such as the shortest route, delay and minimum energy consumption of data transmission~\cite{yiqian.sensors.2013}. Accordingly, we propose a general utility-based value model for market-oriented IoT smart-city information trading. The general form of utility of any system participant can be defined in the form of gross profit, i.e., $\mbox{Utility}\triangleq \mbox{Benefit}-\mbox{Cost}$. Note that in the utility formulation, the benefit can be the gain of the participant in terms of monetary or nominal rewards, depending on the application setting. Moreover, the benefit varies with different system components. Raw data may have a lower value compared with that of higher-level information services provided to users. For example, GPS trace is far cheaper than vehicle routing services for general users. Furthermore, identical raw data can be valued differently by different information processors, depending on their needs and demands. Again, a government agency and commercial map service may value GPS trace differently. Additionally, information received at different time instances leads to considerably diverse values due to timeliness of information. GPS trace which is delayed due to erroneous transmission typically has a low value, especially for real-time traffic prediction.
	
	In this regard, information-centric architecture can well support market-oriented approaches for information trading in smart-city IoT systems. Information value and pricing are affected by the availability, accessibility and quality obtained by different types of users, which are the next topics of discussion.

	\subsection{Incentive, Pricing and Sensing-as-a-Service}
	
	Smart-city applications are complex systems with a variety of participants acting in their own best interests. In the conventional systems, there can be a centralized authority or provider to deploy IoT devices and collect data from them. For smart-city IoT systems, participants are self-interested such as organizations, competitive service providers and human users with various demand. The participants thus need to be incentivized to cooperate and exchange information.
	
	An efficient market-oriented tool to leverage incentives to manage network communication and data resources is named Smart Data Pricing~\cite{sen.sdp.2014}, where dynamic pricing can be applied to network resources. The dynamic pricing approach can both indicate different values of the network resources, e.g., information, as well as exclude or deter some participants which may find the benefit is exceeded by the current price level. Nevertheless, pricing for IoT information exchange in a smart city does not necessarily involve monetary incentives. The participants, especially human users, can be satisfied by different forms of rewards, e.g., priority and social perception. Success stories include crowdsourcing applications such as Foursequare and Pokemon Go, where users are motivated to collect physical world information in exchange of reward points and game achievements.

	Based on the information-centric architecture of smart-city IoT, a promising information trading business framework, namely Sensing-as-a-Service (S2aaS), has been introduced~\cite{perera.saas.2014, guijarro.saas.2017}. The components in a smart city are categorized into four conceptual layers~\cite{perera.saas.2014}, as shown in Fig.~\ref{Fig:Arch}, including: sensors to directly collect data, sensor publishers to transfer the data, service providers to use data and produce services, and users. Participants in the same layer of S2aaS can be potential competitors since they provide complementary or substitutable information and services in the ``IoT market''. Under the S2aaS framework, information can be circulated as a commodity. Additionally, information with different quality will be priced based on their market values in the trading.
	
	\subsection{Information Freshness and Social Impacts}
	
	Information has different nature from traditional physical commodities while being priced. Information has merely no marginal cost while being reproduced, i.e., virtually no cost for copying and forwarding information. By contrast, freshness of information and social impacts affect the value of information more significantly.
	
	\emph{Age-of-Information} (AoI)~\cite{kosta.aoi.2017} is a concept to measure the freshness of information, defined as the time interval from the most recent update of the information at the information recipient side. In the context of smart-city IoT, users may operate as ``free riders'' that wait until the information is disseminated across the IoT network as time elapses. However, as AoI increases with time, the trading value of information decreases immediately after the information is generated. Discriminatory pricing strategies, i.e., based on the time that the information is made available to a participant, can be adopted. As such, the users are required to buy information before the information becomes not fresh or even obsolete.
	
	Equally important, social impacts affect information pricing because of the existence of other participants, defined as \emph{externalities}, which mainly include the network effect and congestion effect. The network effect represents the situation in which some market participants mutually increase their utility when they are in the system. For example, a commuter benefits more from a higher accuracy of travel time when more of other commuters contribute travel information. On the contrary, the congestion effect decreases the utility due to detrimental performance. 
	
	The age-of-information and social impacts directly influence the development of the S2aaS model for smart-city IoT. Specifically, a number of sensor devices increase the impacts on the performance of users. The reason is that service providers may potentially collect information from more sources and with higher precision, which increases the utility of the users in the market.

\section{Design Challenges}	\label{Sec:CaseStudy}
To analyze the current progresses, as well as design issues of market-oriented information trading in smart-city IoT applications, typical cases of market-oriented approaches for smart-city IoT applications are presented in this section.

	\subsection{Raw Data Processing and Trading}
	
	In a smart city, raw data can be generated from time to time. Raw data sensed by sensor devices may have little added value as information. As a result, information processing and trading can be a major business for smart-city participants. Placemeter\footnote{http://www.placemeter.com/} collects and processes video streams of city cameras, turning the video data into meaningful structures information about street situations, e.g., traffic trends and pedestrian directions. Clearly, the processed information has a much higher value than that of raw video data. In this regard, IoT data owners can become micro service providers~\cite{bohli.sigcomm.2009}, as shown in Fig.~\ref{Fig:Arch}, if they implement some video analytics.

	\subsection{Mechanism Design for Resource Allocation}
	In a smart-city IoT application, value of information needs to be revealed to smart city participants. Auctions, as a method for allocating exclusive resources among multiple users, can be employed as a mechanism to valuate information in the smart city for participants by negotiating, asking and bidding deal prices. An auction has been designed for mobile crowdsensing in smart cities~\cite{pouryazdan.crowdsensing.2016}, where an application task runs a reverse auction to select smartphones for data collection. An important advantage of the auction approach is that there are plenty of auction mechanisms being put into practice with desirable economic properties, e.g., market efficiency and social welfare maximization. 

	\subsection{User Competition and Cooperation} 
	Models developed for competition and cooperation can be adopted into smart-city IoT scenarios for physical resource sharing and multiple access control purposes. Participants in smart city can be motivated to cooperate and compete to achieve optimal strategies of resource utilization and data transmission, such as limited system resources, e.g., communication bandwidth, in a distributed fashion. Game-theoretic approaches are usually employed as a market-oriented approach. For example, the Kolkata paise restaurant (KPR) game is proposed in~\cite{park.kolkatagame.2017} to optimally allocate multiple resources among users considering the preferences of the users. Likewise, to improve utility of smart-city IoT users in sharing resources, coalition among the users can be formed~\cite{wu.d2d.2014}.
	
	\subsection{Market-oriented Security in Information Trading} 
	Security issues are important especially for future autonomous information trading processes. Security in an information market involves data protection and secured trading process. For IoT devices, a specially designed module, such as a security auditing module~\cite{ye.audit.2017}, can be implemented to detect and record threats occurred in IoT operations. Furthermore, blockchain technology~\cite{dustdar.value.2016} is introduced to provide protection and storage for distributed transactions, which is inherently suitable for decentralized information trading accounting in smart-city IoT applications~\cite{mulligan.bizmodel.2013}.

Existing works in the literature ignore the social impacts which play an important role in market-oriented approaches and designs for smart-city IoT, as well as correlated trading processes among all the different system levels. This motivates us to introduce a novel game-theoretic model for information trading in smart-city IoT. The model is able to capture competition among the participants together with externality.	
	
\section{A Game-theoretic Model of Information Trading in Smart-city IoT}\label{Sec:ProposedGame}
In this section, we propose a game-theoretic model considering existing externalities in the system. In the model, an IoT service provider offers IoT service to users with a uniform price. Also, the provider pays rewards to IoT information vendors which own sets of IoT devices to generate IoT information.

		\begin{figure}[htbp]
		\begin{center}
		\includegraphics[width=0.80\textwidth]{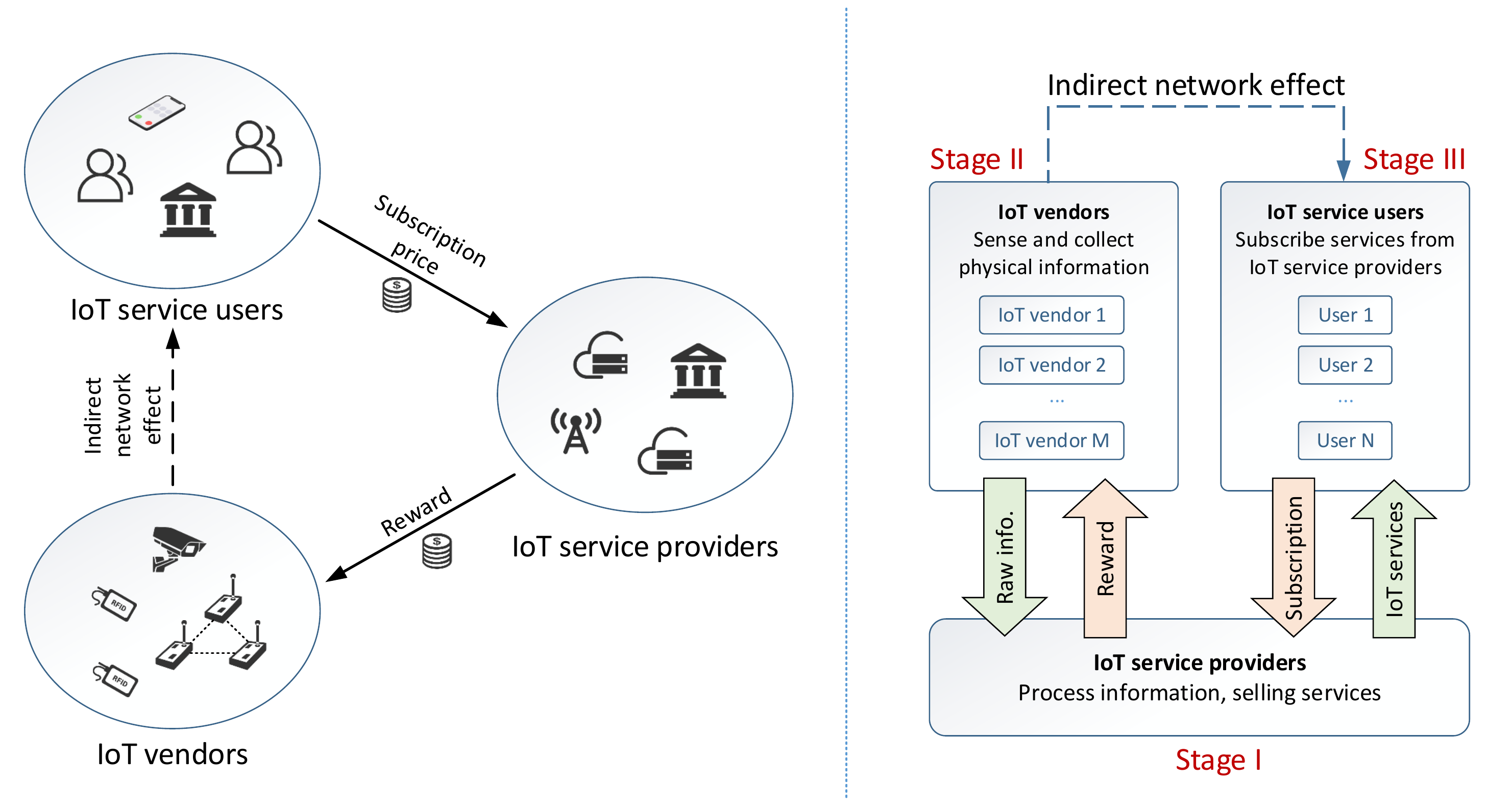}
		\caption{\small System description and the hierarchical sequential game framework for pricing, reward, and IoT service demand modeling.}
		\label{Fig:Sys}
		\end{center}
		\end{figure}
		
	\subsection{System Model}\label{Sec:Model}
	As shown in Fig.~\ref{Fig:Sys}, an IoT service market consists of three types of participants, i.e., game players.
	\begin{itemize}
		\item {\bf IoT service provider}: An IoT service provider works as a resource-rich agent or institution between IoT users and IoT information vendors. The provider allows users to send a demand to purchase IoT services at a fixed subscription price without knowing the details of IoT vendors. The prices are decided by the provider. The provider serves the user demand by obtaining and processing raw IoT information sensed by IoT devices owned by vendors.
		
		\item {\bf IoT information vendors}: A vendor owns a set of different IoT devices to sense and collect raw data. The raw data, e.g., an image from a video camera, is delivered to the provider for further processing for the IoT services. A certain amount of rewards, decided by the vendor, will be charged to the provider.
		
		\item {\bf IoT user}: An IoT service user requires services from the provider in the market. As customer-type participants, the behaviour and decisions of IoT service users can be affected directly and indirectly by other market participants, which is defined as externalities.
	\end{itemize}
	In other words, the vendors, provider, and users are considered as the sellers, resellers, and consumers in the market, respectively.

	We model the interactions among the three players as the three-stage Stackelberg game, with sub-game perfect equilibrium solved to determine the optimal strategies of all the three types of players in the proposed system.
	
	\subsection{Demand Analysis and Utility of IoT Service User}
	In the IoT service market, all the three types of participants aim to optimize their payoff. The payoffs of all the three participants are defined as utility functions.
	
	We consider $N$ IoT users in the market. Each user $i$ determines the demand for IoT services, denoted by $x_i > 0$. The utility $u_i$ can be affected by the following factors:
	\begin{itemize}
		\item Demand level of the user in the market,
		\item Number of vendors accessed by the provider, and hence the number of accessible IoT devices in the system, and
		\item Price charged by the provider.
	\end{itemize}
	Note that when there are more IoT devices, the collected service tends to have better quality from richer available data. Thus, it contributes to the utility of users.
	
	In this case, the benefit of each user $i$ includes:
	\begin{itemize}
		\item \emph{Direct benefit} obtained when user $i$ utilizes the service provided by the provider, denoted as \emph{internal effects}. The benefit is typically modeled as a concave function ${f_i}(x_i)$ of demand that captures the decreasing marginal returns effect, e.g., a linear-quadratic function~\cite{zhang.wcnc.2018}.
		
		\item \emph{Indirect network effects} from the participation level of vendors. Naturally, when more vendors provide sensing data, the quality of IoT service improves, although the user does not directly access the information provided by the vendors.
		
		\item \emph{Direct network effects} caused by all the users in the market. More users in the system lead to additive utility to an individual user due to physical and psychological reasons. For example, a user may trust and be willing to participate in the market, if the user observes that there are many other users including friends already in the network. In this case, the market appears to be attractive to IoT users. The direct network effect term can be expressed as $\xi_{dir}(\mathbf{x})=x_i\sum\nolimits_{j=1}^N g_{ij}x_j$, where $g_{ij}$ is the degree of the network effect that the existence of user $j$ induces the utility of user $i$ and $\mathbf{x}$ contains the demands of all users, i.e., $\mathbf{x}=\{x_1, x_2, \ldots, x_N\}$. 
	\end{itemize}
	The cost incurred to user $i$ (i.e., negative value) includes:
	\begin{itemize}
		\item {\em Direct congestion effects} denoted by $\xi_{con}(\mathbf{x})$ directly affect the user utility negatively. When there are more users with higher demand, congestion happens and service performance is degraded. The congestion effect is typically a convex function in which the marginal cost increases as the total demand increases.
	
		\item Price per unit of IoT service charged to all the users is denoted by $p$.
	\end{itemize}

	\subsection{Utilities of IoT Information Vendor and Service Provider}
	Consider a set of $M$ vendors. Each vendor $j$ decides whether to sell raw sensing data to the provider or not. The utility function of a vendor includes reward (benefit) and cost components. The vendor will participate in the market if reward $r_j$ is higher than cost $c_j$. Thus, the utility of vendor $j$ is denoted by $\Gamma_j =  ( r_j- c_j )\delta_j$. Here, $\delta_j$ is a {\em dispatch demand function}, indicating the amount of user demand that is dispatched/sent directly from the users to vendor $j$. This dispatched demand can happen when the users find that they can use the sensing data from the vendor directly. For example, the IoT devices of the vendor may implement some sensing data processing, e.g., video analytics, which makes IoT information useful and available to the users. The dispatch demand increases when the reward increases and/or the cost decreases. In other words, the vendor serves more demand when it generates more reward or incurs less cost.
	
	The provider decides the unit IoT service prices $p$ charging to users to maximize its utility, i.e., profit, denoted by $\Pi$ which is benefit minus cost. The benefit is generated proportional to the demand and price while the cost is the reward paid to the vendors.

	\subsection{A Hierarchical Stackelberg Game Formulation}

	We adopt a three-stage Stackelberg game theoretic model to analyze the proposed IoT market. Figure~\ref{Fig:Sys} presents the market-oriented information trading among the provider, vendors, and users as a hierarchical Stackelberg game. 
	\begin{itemize}
		\item {\bf Stage I (provider)}: The provider acts as the leader in the game. It decides a price $p^*$ to maximize its utility $\Pi$.
		
		\item {\bf Stage II (vendor)}: Given the optimal provider price $p^*$, Each vendor $j$ decides the reward $r_j$ to be charged to the provider to maximize its utility $\Gamma_j$.
		
		\item {\bf Stage III (user)}: Given the optimal price $p^*$ from the provider and optimal reward $r^*_j$ from the vendor, each user determines the demand $x_i^*$ to maximize its utility $u_i$.
	\end{itemize}
	
	By using backward induction, we derive analytical solutions of the three-stage Stackelberg game. The problem in Stage~III is solved firstly by using the first-order optimality condition. The problems in Stages II and I can be solved in the same manner~\cite{zhang.wcnc.2018}. The solution is the Stackelberg equilibrium which maximizes the utility of the users given the price and rewards of the provider and vendors, respectively. The equilibrium also maximizes the utilities of the vendors and provider. The existence and uniqueness of the Stackelberg equilibrium is proved in~\cite{zhang.wcnc.2018}.

	\subsection{Numerical Results}\label{Sec:Numerical}
	As an example, we evaluate the performance of the IoT market with five users, five vendors, and one provider. We mainly examine the impacts on the utility performance by different system parameters. 

		\begin{figure}
		\begin{center}
		\subfloat[]{\includegraphics[width=0.24\textwidth]{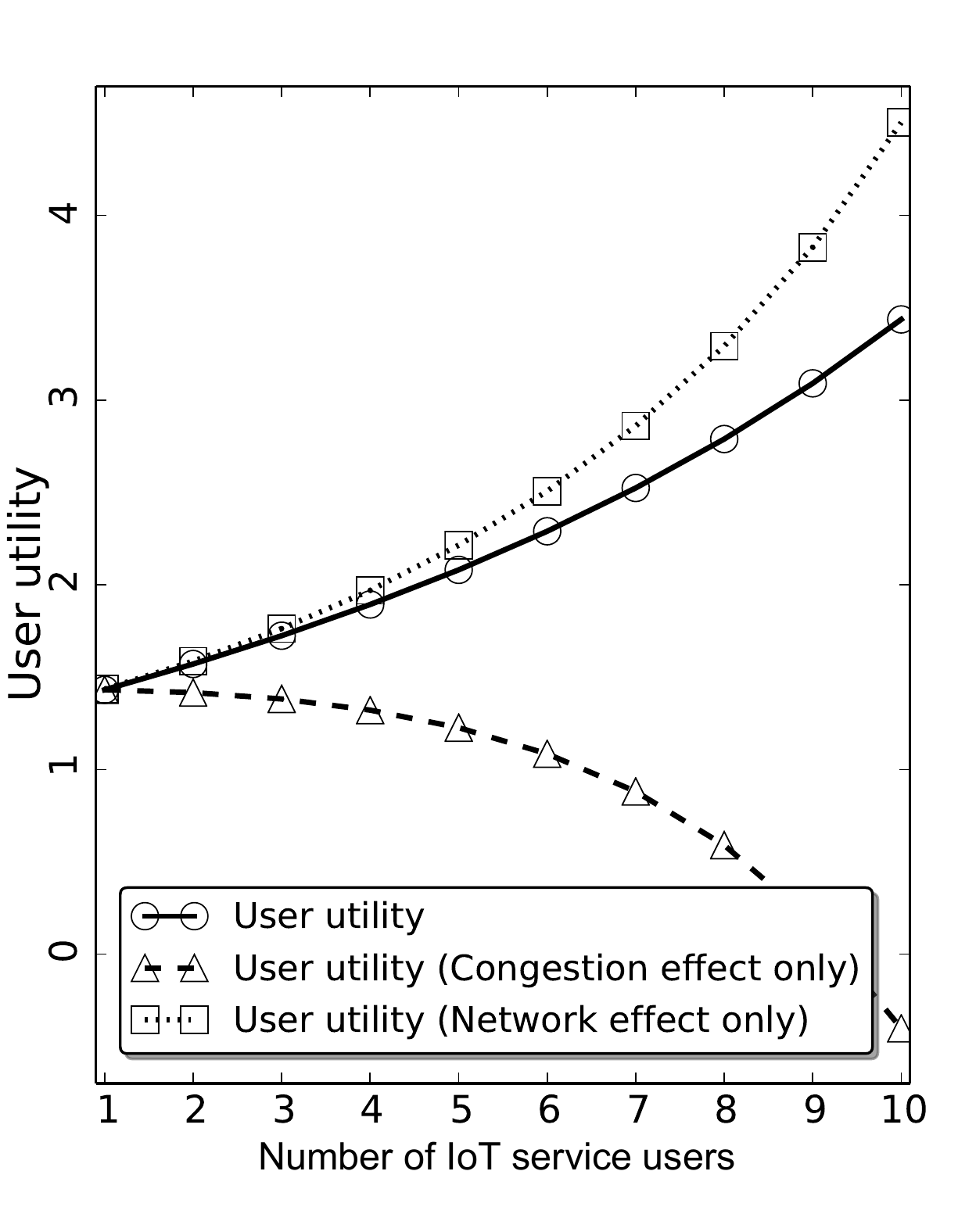}}
		\subfloat[]{\includegraphics[width=0.24\textwidth]{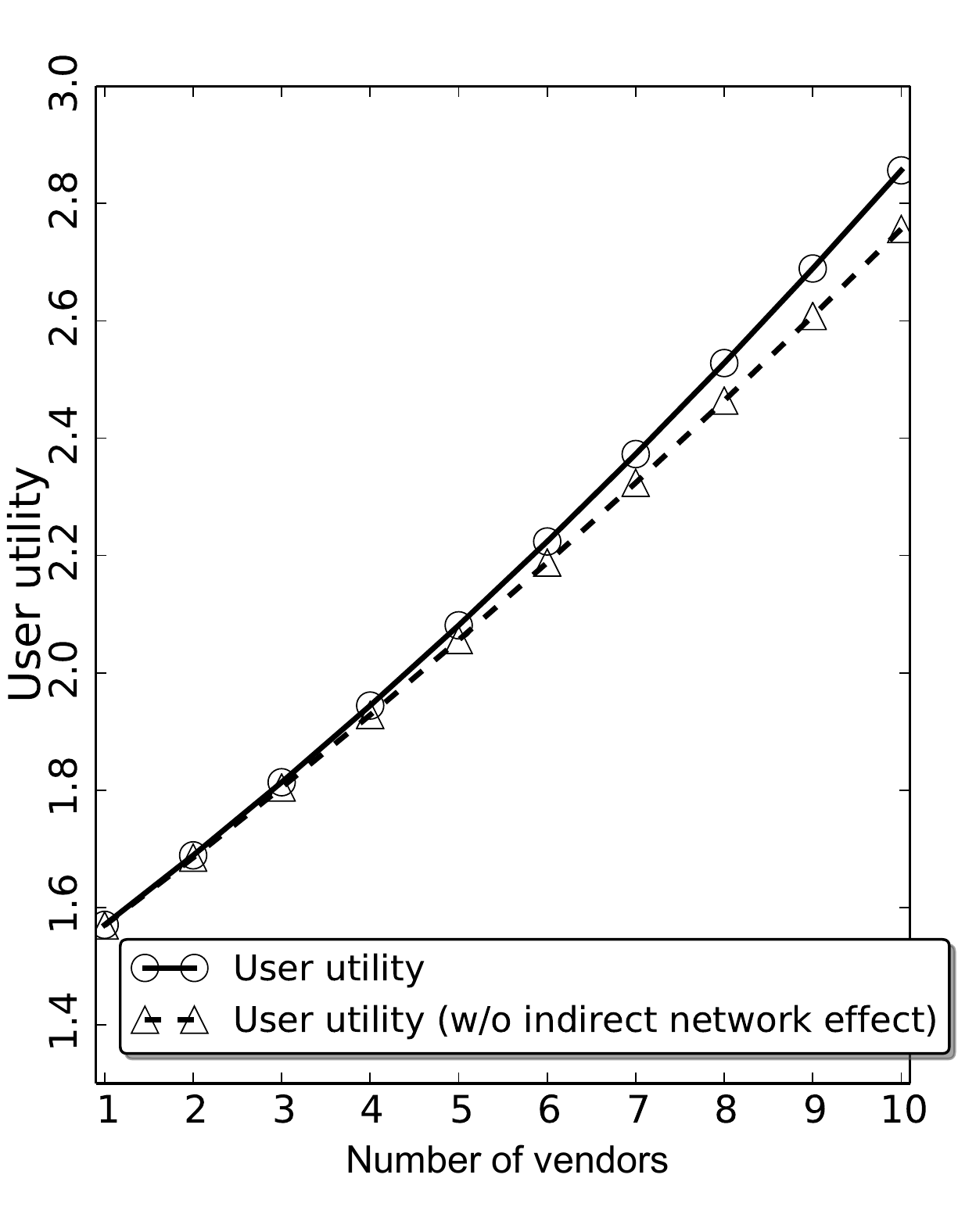}}
		\subfloat[]{\includegraphics[width=0.24\textwidth]{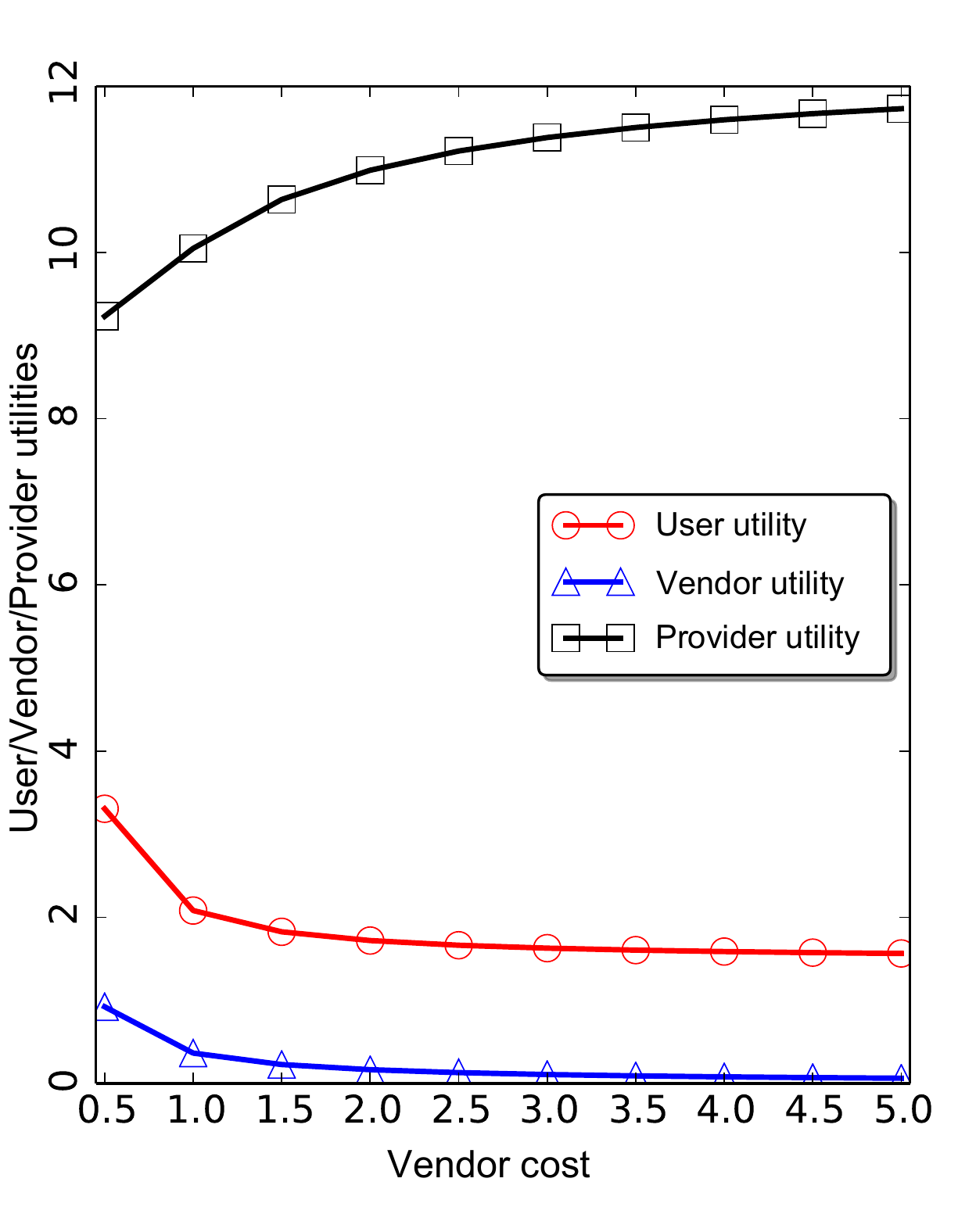}}
		\end{center}
		\vspace{-3mm}
		\caption{\small (a) Impacts of direct network effect and congestion effect on user utility, (b) impacts of indirect network effect on user utility, and (c) impacts of vendor cost on user/vendor/provider utilities.}
		\vspace{-3mm}
		\label{fig:Num.3}
		\end{figure}
		
		We first analyze the impacts of externalities, i.e., direct and indirect network effects, as well as the congestion effect from peer users. As shown in Fig.~\ref{fig:Num.3}(a), network effect increases the user utility. On the contrary, congestion effect leads to lower user utility. In Fig.~\ref{fig:Num.3}(b), indirect network effect introduced by the vendors leads to a higher user utility. That is, the existence of vendors in the system encourages the users to generate more demand.
				
		From Figs.~\ref{fig:Num.3}(a) and (b), direct network effect and congestion effect are both directly correlated with the number of users. When the number of users increases from $1$ to $10$, the impacts become more significant, as shown by the gap of performance curves in Fig.~\ref{fig:Num.3}(a) becoming much wider. By contrast, the indirect network effect caused by vendors has relatively mild impacts on the user utility, as shown in Fig.~\ref{fig:Num.3}(b). When the number of vendors increases, the user utility without indirect network effect is only slightly lower than that when indirect network effect is incorporated. Note that the scenario without indirect network effect happens when the users do not benefit from more vendors participating in the market. For example, the vendors can provide the same sensing data in which more data becomes merely redundant.

		The vendor cost $c_j$ affects the optimal rewards of vendors, and consequently influences the price and user demand. To examine the impacts of vendor cost, we vary the sensing cost of each IoT device owned by the vendors from $0.5$ to $5.0$, as shown in Fig.~\ref{fig:Num.3}(c). This variation can happen, for example, due to different operating IoT device, communication, and processing infrastructure expenses. The vendor utility decreases and approaches $0$ because the cost becomes too high, preventing the user demand from being dispatched to the vendor. In this case, as the vendor cost increases, user demand decreases because of the indirect impact, and the amount of demand to the provider increases.

\section{Conclusion}\label{Sec:Conclusion}
In this paper, general approaches for market-oriented IoT smart cities are studied. With the properties of quantity and heterogeneity of information created and circulated in a smart city, an information-centric architecture of IoT for smart cities has been proposed. Based on the information-centric system architecture, we have discussed the value view of smart city systems for information tradings, with the analyses of incentive for trading, service model, information timeliness and social properties. Finally, an IoT system prototype scenario for information trading has been studied employing Stackelberg game theoretic modeling.


\end{document}